\documentclass[%
 reprint,
 superscriptaddress,
 amsmath,amssymb,
 aps,
prb,
floatfix,
]{revtex4-2}

\usepackage{soul}
\usepackage{xcolor}
\usepackage{graphicx}
\usepackage{dcolumn}
\usepackage{physics}
\usepackage{float}
\usepackage{hhline}
\usepackage{hyperref}

\usepackage{xcolor}

\newcommand{\fakefigure}[1]
{\refstepcounter{figure}\label{#1}}
\newcommand{\faketable}[1]
{\refstepcounter{table}\label{#1}}
\newcommand{\fakesection}[1]
{\refstepcounter{section}\label{#1}}

\begin{document}

\title{Strong localization of microwaves beyond two dimensions in aperiodic Vogel spirals}

\author{Luis A. Razo-L\'opez}
\affiliation{Université Côte d’Azur, CNRS, Institut de Physique de Nice (INPHYNI), France}

\author{Geoffroy J. Aubry}
\affiliation{Université Côte d’Azur, CNRS, Institut de Physique de Nice (INPHYNI), France}
\email{geoffroy.aubry@univ-cotedazur.fr}

\author{Felipe A. Pinheiro}
\affiliation{Instituto de Física, Universidade Federal do Rio de Janeiro, Rio de Janeiro-RJ, 21941-972, Brazil}

\author{Fabrice Mortessagne}
\affiliation{Université Côte d’Azur, CNRS, Institut de Physique de Nice (INPHYNI), France}
\email{fabrice.mortessagne@univ-cotedazur.fr}

\begin{abstract}
Despite extensive studies, strong localization of electromagnetic waves in three dimensions has never been experimentally achieved.
Stepping out of the paradigm of disordered systems, we carry out microwave transport experiments in planar aperiodic Vogel spiral arrays of cylinders with high dielectric permittivity.
We characterize the electromagnetic modal structure in real space showing mixtures of long-lived modes with Gaussian, exponential, and power law spatial decay.
This unique modal structure, which cannot be found in traditional periodic or disordered photonic materials, is shown to be at the origin of strong localization of electromagnetic waves that survives even in a three dimensional environment.

\end{abstract}

\vspace*{0.9em}

\maketitle

\section{Introduction}

A full understanding and control over electromagnetic transport in photonic media is crucial for the efficient design of optical structures, paving the way for many applications~\cite{vynck2021light}.
Controlling light transport in photonic structures involves the ability not only to understand, but also to engineer the electromagnetic modes that such structures can support.
In the case of traditional periodic and disordered optical structures, light transport and the underlying electromagnetic modal structure have been extensively investigated over the years.
In periodic photonic structures the scattering of propagating, extended electromagnetic waves from Bragg planes is related to the opening of photonic bandgaps at certain frequencies~\cite{joannopoulos2008molding}.
In disordered optical systems the interference of multiply scattered waves may lead to the formation of exponentially localized states and eventually to the breakdown of light diffusion~\cite{lagendijk2009fifty,Schertel2019b,Laurent_2007}. 
This effect, the optical counterpart of Anderson localization for electrons in solids~\cite{Anderson1958}, strongly depends on dimensionality~\cite{Abrahams1979}, and, in three-dimensional (3D) dielectric media, there is no unquestionable observation of light localization transition so far for disordered systems~\cite{SkipetrovPRL,Sperling,SkipetrovNJP,Yamilov2023}.
In two-dimensional (2D) media, when structural correlations are introduced, as it is the case of hyperuniform disordered materials~\cite{torquato2003local}, a richer transport diagram exists that include transparency, light diffusion, Anderson localization, or full band gaps, depending on the frequency~\cite{leseur2016high,ricouvier2017optimizing,froufe2016role,froufe2017band,ricouvier2017optimizing,haberko2020transition,aubry2020experimental, Klatt2022, Monsarrat2022, FroufePerez2023, Granchi2023}.
 
As an alternative to periodic and disordered photonic structures, aperiodic metamaterials designed by means of deterministic mathematical rules have emerged as a novel material platform for photonic devices~\cite{dal2022waves,DalNegroReview,DalNegroCrystals}.
Indeed, these structures exhibit unique optical properties that do not exist in either periodic or disordered photonic media, such as fractal transmission spectra~\cite{sgrignuoli2020multifractality,reisner2022experimental}, subdiffusive transport~\cite{sgrignuoli2020subdiffusive}, and light localization transition~\cite{sgrignuoli2019localization}. 
From a technological point of view, these unusual optical properties have fostered the development of functionalities that also cannot be found in conventional periodic or disordered structures, including applications in lasing~\cite{noh2011lasing,Vardeny2013}, optical sensing~\cite{razi2019optimization,Lee,Gopinath}, photo-detection~\cite{trevino2011circularly}, and optical imaging~\cite{huang2007optical}.
Among various classes of deterministic aperiodic photonic media, Vogel spiral arrays single out for its versatility and the possibility to tailor its structural order~\cite{dal2012analytical,christofi2016probing,Pollard} and light-matter interactions~\cite{trojak2020cavity,trojak2021cavity}.
The unusual optical properties of aperiodic systems are enabled by their unique electromagnetic modal structure.
For instance, recently it was theoretically demonstrated that aperiodic Vogel spiral arrays display a rich spectrum of long-lived and spatially localized quasimodes with distinctive spatial decay forms, namely Gaussian, exponential, and power law~\cite{prado2021structural}.

In the present article, we not only experimentally demonstrate that these characteristic types of electromagnetic modes coexist in Vogel spirals, but also that this unique electromagnetic modal structure leads to unusual wave transport phenomena.
Indeed, by conducting microwave transport experiments in Vogel spiral arrays of cylinders with high dielectric permittivity, we unveil the consequences of this peculiar modal structure on wave transport and localization.
In particular, we show that the presence of long-lived quasimodes with exponential, power law and Gaussian spatial decays is at the origin of a very slow decay of the electromagnetic energy that propagates throughout the arrays.
Furthermore, we experimentally demonstrate that these localized modes satisfy the Thouless criterion for Anderson localization in random systems, despite the fact that the propagation medium is not disordered, and that the modes are not necessarily exponentially localized in space.
Finally, and this is the main result of our paper, we demonstrate the robustness of these long-lived modes against the change in the dimensionality of the cavity, beyond the 2D limit, preserving their spatial profiles and quality factors even when the homogeneity of the electric field in the $z$-direction is broken.
Strong localization of light in 3D is a long standing, sought-after goal in the field of waves in complex media~\cite{Sperling} that we experimentally achieve not with uncorrelated disordered structures, as it has been unsuccessfully tried for many years, but rather with deterministic aperiodic systems.

\section{Experimental Setup}

\begin{figure}
    \centering
    \includegraphics{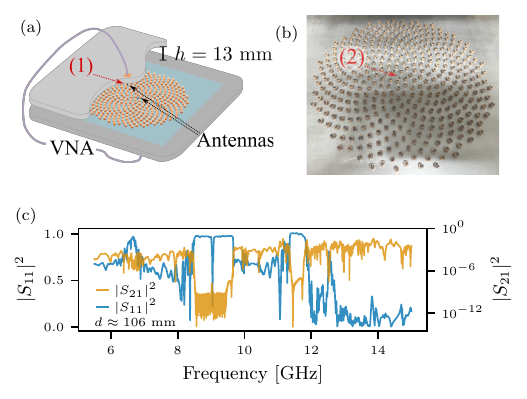}
    \caption{
    (a) Typical configuration of the experimental microwave setup. Cylinders are placed between two parallel aluminum plates separated by a distance $h=13$~mm. A fixed antenna (2) is placed at the origin of the reference system $(x,y)=(0,0)$ while an other antenna (1) is positioned in the center of the movable top plate $(x,y)$.
    (b) Image of the 2D array of dielectric cylinders. The top plate has been removed to reveal the details of the sample.
    (c) Reflected $|S_{11}(\nu)|^2$ (blue line) and transmitted coefficients  $|S_{21}(\nu)|^2$ (orange line) for the GA spiral at a given distance $d=\sqrt{x^2+y^2}$ ($x=75$~mm, $y=75$~mm) from the origin of the reference system.
    }
    \label{fig:Setup}
\end{figure}

Our main sample consists of $N=390$ cylindrical scatterers (dielectric permittivity $\varepsilon\simeq45$, radius 3~mm and height 5~mm) disposed in a cavity made of two parallel aluminium plates which are separated by a distance $h=13$~mm [see Fig.~\ref{fig:Setup}(a)].
Cylinders are placed following a Vogel spiral array of 140~mm radius with a planar density $\rho\approx0.65$~cm$^{-2}$ [see Fig.~\ref{fig:Setup}(b)] and each of them is a Mie scatterer~\cite{aubry2020experimental}.
This lattice, often called \emph{Golden-Angle spiral} (GA spiral), is defined in polar coordinates $(r,\theta)$, as $r_n=a_0\sqrt{n}$ ($a_0=6.93$~mm) and $\theta_n$ linked to the golden number (see Appendix~\ref{sec:VogelSpirals}).
In order to ensure an homogeneous electrical contact between the scatterers and the bottom plate and thus a good reproducibility, we covered the bottom plate with a self-adhesive thin plastic film.

As sketched in Fig.~\ref{fig:Setup}(a), the electric field in the cavity is mapped by the straight antenna (1) placed at the center of the movable top plate.
We measure all points on a $5\times5$~mm$^2$ grid, covering the disk region occupied by the spiral plus one corner of the embedding square [see, e.g., Fig.~\ref{fig:RadialDecay}(a)], resulting in $\sim 3600$ measured points (see Appendix~\ref{sec:ExpDet}).
\begin{figure}
    \centering
    \includegraphics{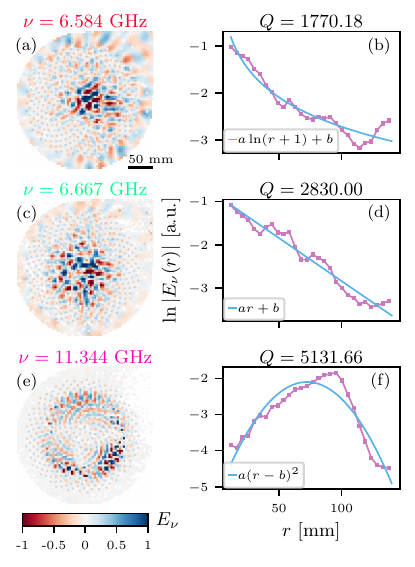}
    \caption{Spatial modal structure and radial profile of representative eigenmodes with characteristic 
    (a)-(b) power law,
    (c)-(d) exponential, and
    (e)-(f) Gaussian decay.
    Amplitude maps are normalized such that $\max(\abs{E_\nu})=1$.
    Radial decays (pink dots) are obtained by performing an azimuthal average operation where $r=\sqrt{(x+5)^2+y^2}$ {\it{i.e.}}, $r$ is measured from the geometrical center of the spiral and expressed in mm. Different radial decays are discriminated by minimizing the sum of squared residuals.
    }
    \label{fig:RadialDecay}
\end{figure}
The presence of the second straight antenna~(2) at the center of the bottom plate defines the origin of the system coordinates $(x,y)=(0,0)$ and allows to measure both the complex reflection and transmission signals $S_{11}(\nu)$ and $S_{21}(\nu)$, respectively, using a Vector Network Analyzer (VNA).
Note that both antennas are linear and perpendicular to the plane of the cavity, thus imposing a transverse magnetic (TM) polarization (electric field perpendicular to the plane of the cavity).
For a height $h=13$~mm, the empty cavity can be completely considered as 2D below the cutoff frequency $\nu_{\mathrm{cut}}=c_0/\left(2h\right)\approx11.5$~GHz, where $c_0$ is the speed of light in air. 
Beneath this threshold, just the fundamental transverse magnetic mode TM$_0$ can propagate in air and the field is invariant along the $z$-axis.
Measured positions are scanned in a frequency range between 5.5 and 15~GHz meaning that both 2D and 3D regimes can be investigated.
Examples of the measured spectra $|S_{21}|^2$ as well as $|S_{11}|^2$ are shown in Fig.~\ref{fig:Setup}(c) for a given distance $d$ between both antennas, where the geometrical center of the spiral is placed at $(x,y)=(-5\,\mathrm{mm},0)$.
Vanishing transmission values (reflection values close to~1) at certain frequencies indicate the presence of bandgaps~\cite{TrevinoMF}.
Outside of these gaps, the transmitted signal is a superposition of peaks which are related with the resonances of the system. 
The parameters of each resonance (frequency $\nu$, width $\delta \nu$ and complex amplitude) are extracted by means of the harmonic inversion technique as described in Refs.~\cite{JMain_2000,Wiersig_2008}.
Later, by clustering the amplitudes of the same resonance measured at all positions, the map of the electric field amplitude $E_\nu(x,y)$ is obtained revealing the spatial structures of each eigenmode of the system~\cite{aubry2020experimental}.

\section{Conductance in Vogel Spiral}

Figure~\ref{fig:RadialDecay} shows the spatial modal structure of three characteristic eigenstates found in the same experimental GA spiral and their corresponding radial decay.
Specifically, power-law, exponential, and Gaussian radial decays have been identified by minimizing the sum of squared residuals.
Our experimental results demonstrate that Vogel spirals support a rich variety of long-lived modes that exhibit different spatial extent and radial decay profiles, hence confirming recent theoretical predictions~\cite{prado2021structural}.
This can be contrasted to disordered samples, where (Anderson) localized states are always characterized by an exponential radial decay (see Appendix~\ref{sec:DisSys}).
The analyzed long-lived modes found experimentally in this GA spiral sample are distributed in three frequency windows around $\nu\sim6.6$~GHz (exponential, power-law and Gaussian modes), $\nu\sim8.3$~GHz and $\nu\sim11.2$~GHz (Gaussian modes) and are characterized by high quality factors $Q=\nu/\delta\nu$, i.e., low energy-loss ratios.

\begin{figure}
    \centering
    \includegraphics{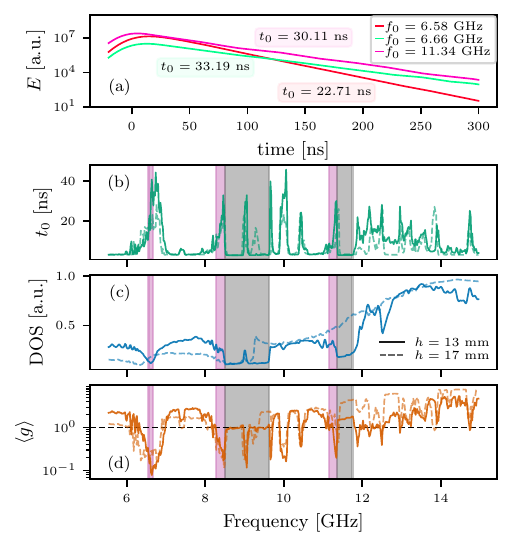}
    \caption{
    (a) Evolution of the energy as function of time for three different frequency centers $f_0$ and $\Delta\omega=0.01$~GHz.
    (b) Characteristic decay time $t_0$ as a function of the frequency. The total frequency range have been mapped by 472 frequency filters spaced by $\Delta f=0.02$~GHz with $\Delta\omega=0.01$~GHz.
    (c) Experimental density of states (DOS).
    (d) Experimental Thouless conductance $\expval{g}$ as a function of the frequency. The dotted line indicates $\expval{g}=1$.
    Frequency windows with the analyzed long-lived states (exponential, power-law or Gaussian) or band gaps have been highlighted (violet or grey stripes, respectively).
    Solid [dashed] lines in (b), (c) and (d) correspond to the case $h=13$~mm [$h=17$~mm].
    }
    \label{fig:Transport}
\end{figure}

Dynamical electromagnetic transport properties can be probed by measuring the temporal evolution of the energy carried by a certain superposition of modes by means of the transmission spectra as $E=\sum_\textrm{all positions} \left|\mathcal{F}\left\{S_{21}\times F_{f_0,\Delta\omega}(\nu)\right\}\right|^2$, where $\mathcal{F}\left\{\cdot\right\}$ represents the Fourier-transform and $F_{f_0,\Delta\omega}(\nu)$ a Gaussian band pass filter of bandwidth $\Delta\omega$ centered around $f_0$. 
Figure~\ref{fig:Transport}(a) displays the energy as a function of time for three different filters centered around the frequency of the modes extracted in Fig.~\ref{fig:RadialDecay} and with $\Delta\omega = 0.01$~GHz.
Within these frequency intervals, the presence of modes with high quality factors leads to very slow energy dynamics.
Assuming an exponential decay of the energy with respect to time, $E\sim \exp\left(-t/t_0\right)$, one can fit a characteristic decay time $t_0$ that is closely related to the average width of the modes contributing to the transport $\expval{\delta\nu}\sim 1/t_0$.
Next, by repeating the previous analysis in a systematic way, we compute the characteristic decay time $t_0$ as a function of the frequency [see Fig.~\ref{fig:Transport}(b)].
We focus on frequency ranges where the spatial structure of the eigenmodes can be properly characterized by the harmonic inversion/clustering methods.
Here, $t_0$-regions associated with frequency windows containing exponential, power-law or Gaussian states (pink stripes) are characterized by high peaks whose maximum values are of the order of those found in disordered systems (see Appendix~\ref{sec:DisSys}), whilst $t_0$-valleys correspond to short-lived, not spatially localized states or band gaps.

The existence of band gaps can be investigated by experimentally extracting the density of states (DOS) that is directly accessible from the intensity of the reflected signal as $\mathrm{DOS}\approx 1-\expval{\left|S_{11}\right|^2}_\mathrm{all \, positions}$~\cite{Kuhl2010,Barkhofen2013}.
In Fig.~\ref{fig:Transport}(c), we plot the DOS, where two band gaps are observed (grey stripes).
The two lower band edges are populated by Gaussian states~\cite{liew2011localized}, while no modes are found by the harmonic inversion/clustering methods in the band gaps, except for the peaks at the center of the first band gap that are a signature of defect modes of the system.
Additionally, an increment in the number of states supported by the system is observed above $\nu_{\mathrm{cut}}$.

The Thouless conductance, defined as $g=\delta\nu/\Delta\nu$ (where $\Delta\nu$ is the spacing between consecutive resonances), is a key quantity in localization theory and it is used as a fundamental criterion for Anderson localization in disordered systems.
Indeed, the Thouless criterion establishes that Anderson localization occurs for $g<1$~\cite{Thouless_1997,Wang2011,Mondal_2019}.
Using the two previously introduced quantities (DOS and $t_0$), we extract experimentally the average Thouless conductance [see Fig.~\ref{fig:Transport}(d)] as $\expval{g}\sim \mathrm{\expval{DOS}}_{\Delta f}/(t_0 \Delta f)$ where $\expval{\mathrm{DOS}}_{\Delta f}$ is the average density of states over the frequency interval $\Delta f$.
Figure~\ref{fig:Transport}(d) experimentally demonstrates that non-random systems can fulfil the Thouless criterion for Anderson localization, originally conceived to characterize localization in disordered structures, again confirming theoretical predictions~\cite{sgrignuoli2019localization}. 
Our findings confirm that not only eigenmodes characterized by an exponential spatial decay can satisfy the Thouless condition, as it occurs for disordered systems, but also other modes with different spatial decay forms, such as algebraic and Gaussian decays.  
To the best of our knowledge localized, long-lived modes that fulfil the Thouless criterion in non-random arrays with non-exponential spatial decays have never been experimentally observed so far. This result demonstrates experimentally the unique modal structure that aperiodic Vogel spirals support, leading to unusual properties of wave transport and localization.
The existence of long-lived modes with different decay types in the same system means that these distinct classes of modes will have different sensitivity to the sample boundaries.
As a result, these classes of modes exhibit different evolution of the transport quantities with respect to the system size, as shown in the following section.

\section{Scaling Analysis} \label{sec:ScaAna}

\begin{figure}
    \centering
    \scalebox{.97}{\includegraphics{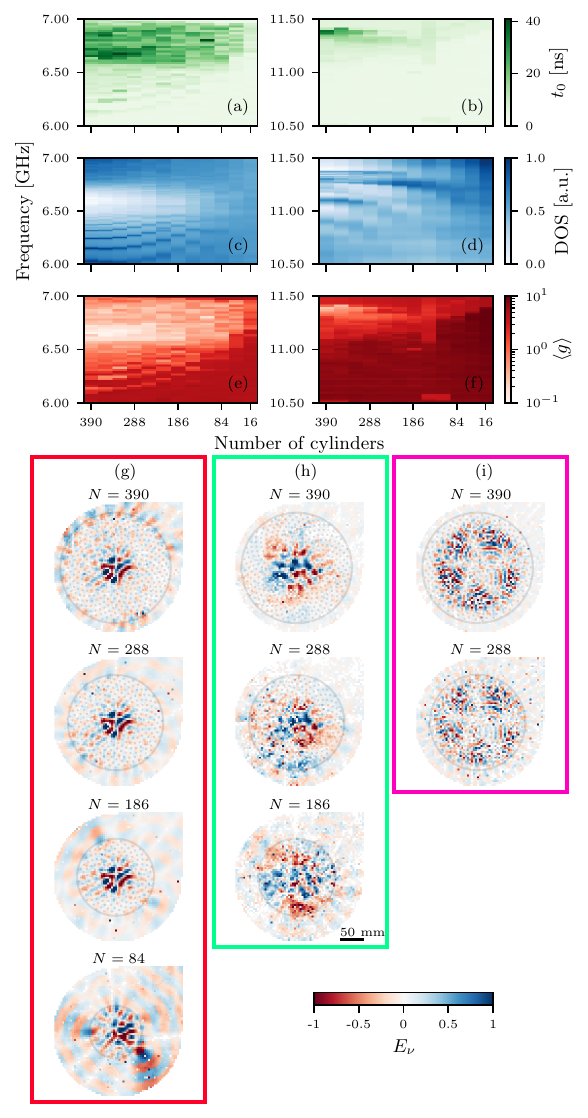}}
    \caption{
    (a)-(b) Characteristic decay time $t_0$ as a function of the frequency and of the number of cylinders. The total frequency range has been mapped by 47 frequency filters spaced by $\Delta f=0.02$~GHz with $\Delta\omega=0.01$~GHz.
    (c)-(d) Experimental normalized density of states (DOS).
    (e)-(f) Experimental Thouless conductance $\expval{g}$.
    Spatial modal structure as a function of the number of cylinders of modes with (g) power-law ($\nu=6.586$~GHz), (h) exponential ($\nu=6.646$~GHz), and (i) Gaussian ($\nu=11.357$~GHz) radial decay.
    The grey circles show the boundary of the samples.
    }
    \label{fig:Size}
\end{figure}

To investigate how the transport quantities as well as the modes are affected by the boundaries of the spiral, the experiment is repeated for 12 different configurations, and for each one of them, the number $N$ of cylinders in the array is reduced according to $N=390-34\times i$ ($i=0,1,\cdots,11$).
In Fig.~\ref{fig:Size}, we show [(a) and (b)] the characteristic decay time $t_0$, [(c) and (d)] the normalized density of states, and [(e) and (f)] the average Thouless conductance $\expval{g}$ as a function of the number of cylinders and frequency in two frequency windows containing the analyzed long-lived modes (from 6 to 7~GHz and from 10.5 to 11.5~GHz).
The spatial structure of three different eigenmodes with characteristic (g) power-law, (h) exponential, and (i) Gaussian decays found by the harmonic inversion/clustering methods are also shown in Fig.~\ref{fig:Size}.
Note that below a certain threshold $N$ [(g) $N\simeq84$, (h) $N\simeq186$, (i) $N\simeq288$], the number of resonances recovered by the harmonic inversion is found to be insufficient to form a cluster and subsequently to reveal the spatial structure of the modes.
Around $\sim 6.6$~GHz, Fig.~\ref{fig:Size}(a) and (e) show that the localization signatures earlier analyzed (high $t_0$ values and $\expval{g}<1$) remain unperturbed even for lattices with a reduced number of cylinders $N\simeq84$, while Fig.~\ref{fig:Size}(c) shows that the density of states start to lose its structure and becomes flat below $N\simeq152$. 
Here the leaking of the wave out of the spiral is driven by the absence of certain cylinders needed to support the long-lived modes.
This fact is illustrated by the power-law and exponential modes [see Fig.~\ref{fig:Size}(g) and (h), respectively] which cannot be recovered by our analysis whenever the typical system size becomes smaller than the noticeable modal size when $N\simeq84$ and $N\simeq186$, respectively.
At higher frequency, Gaussian modes are found to have larger sizes, and are therefore more sensitive to changes in the system boundaries so that they can only exist for larger systems $N\geq288$.
This value of $N$ also determines a critical value above which Gaussian modes with high $t_0$ ($\expval{g}<1$) values disappear [Fig.~\ref{fig:Size}(b)].
Nevertheless, the DOS remains unperturbed even for smaller systems ($N\simeq186$).

\section{2D/3D Transition}

It should be noted that, up to this point, all long-lived modes shown have been found below the 2D cutoff frequency in air ($\nu_{\mathrm{cut}}\approx11.5$~GHz), so the electromagnetic field is confined in the plane of the array.
In order to study the robustness of GA spiral modes with respect to the dimensionality, the distance between both aluminium plates is increased from $h=13$~mm to $h=17$~mm implying a new cutoff frequency $\nu_{\mathrm{cut}}\approx 8.8$~GHz.
As a result the electromagnetic field is actually three-dimensional beyond this frequency.
Then previous experimental procedure and data analysis are repeated, thus the characteristic decay time $t_0$, the DOS, and the average Thouless conductance $\expval{g}$ are extracted and correspond to the dashed lines depicted in Fig.~\ref{fig:Transport}(b), (c) and (d), respectively.
$t_0$ displays remarkable similarities in both cases, being just differentiated by a decrease of the characteristic peaks related to the first localization region (at $\sim 6.6$~GHz) and the formation of a new peak in the first band gap.
The appearance of this peak can be also observed in the DOS, where the size of the first band gap has been considerably reduced in its upper part.
Additionally, the presence of new modes at lower frequencies leads to the population of the second band gap that has completely disappeared, confirming their 3D character.
Nevertheless, $\expval{g}$ still drops by around one order of magnitude near $\sim11.2$~GHz, keeping its shape in all frequency windows containing long-lived modes.

\begin{figure}
    \centering
    \includegraphics{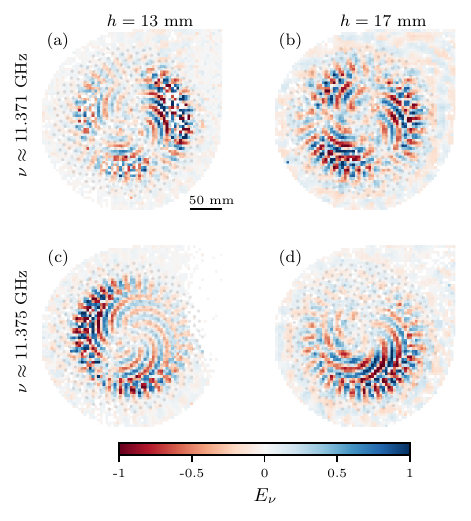}
    \caption{
    Spatial modal structure of two Gaussian long-lived modes at $\sim11.2$~GHz with frequencies [quality factors] (a)~$\nu=11.346$~GHz [$Q=5294$], (b)~$\nu=11.396$~GHz [$Q=3899$], (c)~$\nu=11.35$~GHz [$Q=5170$] and (d)~$\nu=11.401$~GHz [$Q=4159$] and a distance between plates (a)-(c) $h=13$~mm and (b)-(d) $h=17$~mm.
    Amplitude maps are normalized such that $\max(\abs{E_\nu})=1$.
    }
    \label{fig:Dimension}
\end{figure}

In the last part of this work we focus in the highest-frequency window in which long-lived exist ($\sim11.2$~GHz), and which occurs beyond the new cutoff frequency.
Here the presence of long-lived modes previously predicted by the analysis of $t_0$ and $\expval{g}$ is verified by the Harmonic inversion/clustering methods.
Four different Gaussian modes corresponding to this frequency band are shown in Fig.~\ref{fig:Dimension} for both cavity sizes [(a)-(c) $h=13$~mm and (b)-(d) $h=17$~mm].
High quality factors confirm the existence of long-lived modes even in the case where the 2D confinement of the electromagnetic field is not due to the geometry of the cavity.
Indeed, Fig.~\ref{fig:Dimension} experimentally proves the robustness of Gaussian long-lived modes in Vogel spirals against the situation in which the electric field is inhomogeneous in the $z$ direction.
This result also experimentally confirms previous numerical findings~\cite{sgrignuoli2019localization} that demonstrate the existence of localized, long-lived 3D electromagnetic modes supported by 2D Vogel spirals arrays precisely for $\rho c_0^2/\nu^2>3.5$, corresponding to $\nu<12.9$~GHz in our experimental system.

\section{Conclusion}

In conclusion, we have experimentally revealed the spatial modal structure supported by quasi-two-dimensional arrays of dielectric cylinders placed according to aperiodic Vogel spirals, and its impact in wave transport.
We showed that these lattices support a unique modal structure where long-lived modes with different radial decay types (exponential, power-law and Gaussian) coexist, confirming recent theoretical predictions~\cite{prado2021structural}. 
We also investigate the impact of these peculiar modal structure on microwave transport properties by means of the temporal evolution of the energy as well as the characteristic decay time $t_0$ (related with the average resonance width $\expval{\delta \nu}$), the density of states, and the Thouless conductance.
This analysis reveals that Vogel spirals exhibit very slow energy dynamics.
Indeed, we show that in frequency windows containing long-lived modes the values of $t_0$ are similar to those found in the Anderson localized regime in traditional 2D disordered systems.
In these same frequency windows the Thouless criterion for Anderson localization is shown to be fulfilled despite the lack of disorder and the presence of non-exponentially localized modes. 
Long-lived modes supported by Vogel spiral are also proven to be robust against the change in the dimensionality of the cavity, from 2D to 3D, so that they remain essentially unperturbed and preserve their spatial profiles and quality factors even when the homogeneity of the electric field in the $z$-direction is broken.
Thus, without  disorder, our results embark on the long-standing quest for Anderson localization of light in 3D.

\section*{Acknowledgements}
We thank L. Dal Negro, F. Sgrignuoli, M. Prado, and Y. Chen for enlightening discussions. F.A.P. acknowledge CNPq, CAPES, and FAPERJ for financial support. L.A.R.-L. gratefully acknowledges the financial support from CONACyT (Mexico), through the Grant No.775585, and from the French government, through the UCA\textsuperscript{JEDI} Investments in the Future project managed by the National Research Agency (ANR) with the reference number ANR-15-IDEX-0001.

\appendix

\section{Definition of Vogel spirals}
\label{sec:VogelSpirals}

Vogel spirals are defined by their polar coordinates $(r,\theta)$ as
\begin{equation}
    \begin{split}
        r_n & =a_0\sqrt{n}, \\
        \theta_n & = n\alpha,
    \end{split}
\end{equation}
where $n=1,2,\dots$ is an integer, $a_0$ is a positive constant, and $\alpha$ is an irrational number. 
The scaling factor $a_0$ sets the particle separation while $\alpha$ is the divergence angle and determines the constant aperture between successive point. 
The angle $\alpha$ is specified as a function of the irrational number $\xi$ as $\alpha=2\pi\left[1-\mathrm{frac}(\xi)\right]$ where $\mathrm{frac}(\xi)$ is the fractional part of $\xi$.
When $\alpha$ is irrational, point patterns are characterized by a lack of both translational and rotational symmetries.

\begin{figure}
    \includegraphics{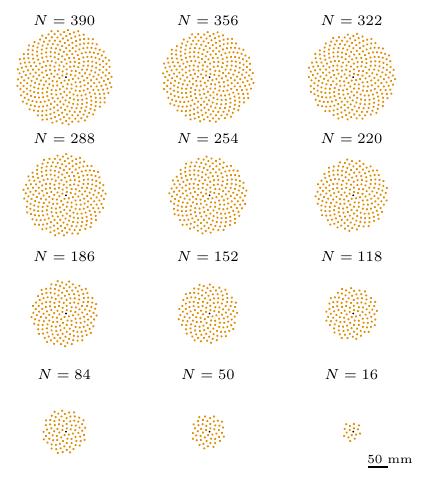}
    \caption{
    Golden-Angle spiral arrays consisting of $N$ cylinders created with $a_0=6.93$~mm and $\xi=(1+\sqrt{5})/2$. The black dot sets the position of the fixed antenna with respect to the cylinder pattern.
    }
    \label{fig:Spiral}
\end{figure}

In this work, we focus in the Golden-Angle (GA) Vogel spiral, also known as ``sunflower spiral'', which is obtained by considering $\xi$ as the golden number $\xi=(1+\sqrt{5})/2$ leading to $\alpha\approx 2.4$ $(137.508^\circ)$ also called the ``golden angle''. 
Experimentally, 12 samples consisting of $N=390-34\times i$ ($i=0,1,\cdots,11$) cylinders are placed and measured for different heights of the cavity ($h=13$ and 17~mm are presented in this work).
All samples are characterized by the same scaling factor $a_0=6.93$~mm and their centers are always located at $(x,y)=(-5~\mathrm{mm},0)$ where the central antenna sets the origin $(x,y)=(0,0)$. 
Figure~\ref{fig:Spiral} shows the 12 different GS experimentally studied in this work.

\section{Experimental map}
\label{sec:ExpDet}

\begin{figure}
    \includegraphics{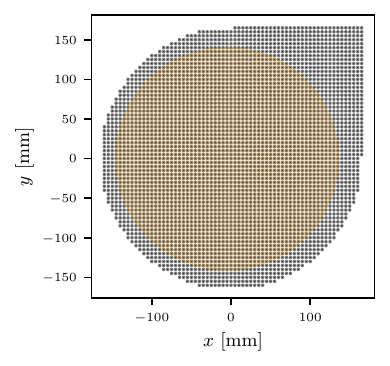}
    \caption{
    Experimental map used to scan the cavity (grey dots). Orange circle indicates the position of the array of cylinders (GA spiral with $N=390$).
    }
    \label{fig:ExpMap}
\end{figure}

Experimental measurements are carried out over a surface determined by a circle of 160~mm radius centred in the origin $(x,y)=(0,0)$. 
In order to observe the electric field out of the lattice, we superimpose a 165~mm size square with corners at $(x,y)=(0,0)$, $(x,y)=(165~\mathrm{mm},0)$, $(x,y)=(0,165~\mathrm{mm})$ and  $(x,y)=(165~\mathrm{mm},165~\mathrm{mm})$.
The resulting area is mapped in a regular $5\times5$~mm$^2$ grid unit cell.
Figure~\ref{fig:ExpMap} shows the experimental map used to scan the cavity where each grey point represents a measured point and the orange circle the space occupied for the array of cylinders ($N=390$).
The total of measured positions is $n=3675$.

\section{Disordered system}
\label{sec:DisSys}

\begin{figure}
    \centering
    \includegraphics[width=\columnwidth]{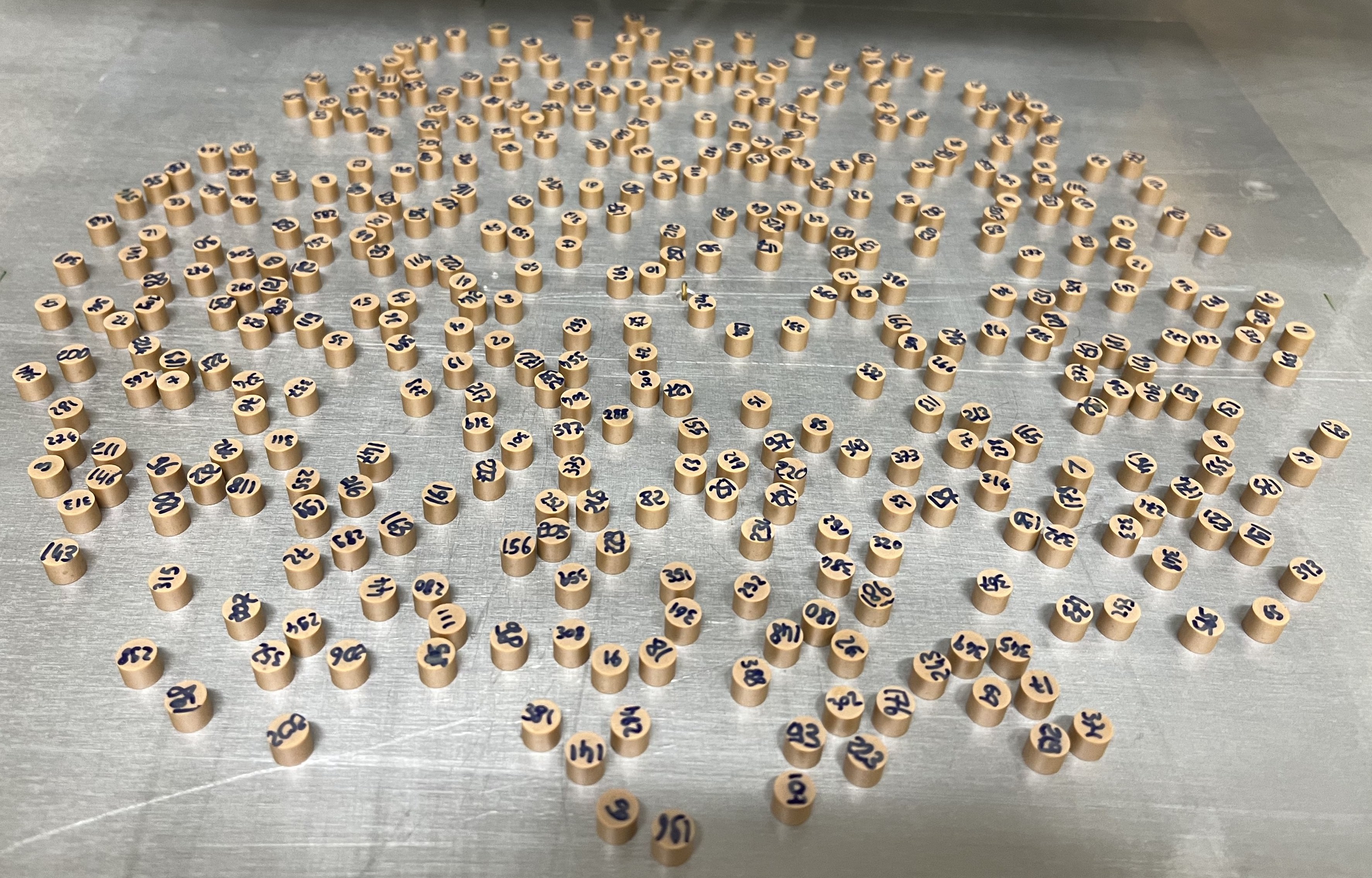}
    \caption{
    Image of the 2D disordered array of dielectric cylinders. The top place has been removed to reveal the details of the sample.
    }
    \label{fig:Dis_lattice}
\end{figure}

For the sake of comparison, besides the main analysis carried out in the GA spiral, we also study the modal structure and microwave transport in a traditional disordered systems (DS).
The disordered point pattern is generated using the software developed in \cite{froufe2016role} and considers a set of $N=388$ packing hard disks of radius $R=3.25$~mm enclosed into a circle of radius 140 mm, thus the planar density is constant respect to the GA spiral case ($\rho\approx0.65$~cm$^{-2}$, see Fig.~\ref{fig:Dis_lattice}).
Here, $R=3.25$~mm is the size of our experimental cylinders plus the size of the tube used to place the cylinders.
The cavity is mapped using the same experimental map (see Fig.~\ref{fig:ExpMap}) in the same frequency range (from 5.5~GHz to 15~GHz) than the GA spiral case.
Both reflected and transmitted signals are measured.
Data analysis to obtain the eigenmodes of the systems (via the harmonic inversion/clustering method), the characteristic decay time $t_0$ (via the Fourier transform), the Density of States (via the reflected signal) and the Thouless conductance $\expval{g}$ (via $t_0$ and the $\expval{\mathrm{DOS}}$) is performed as explained in the main text.

\begin{figure}
    \centering
    \scalebox{.97}{\includegraphics{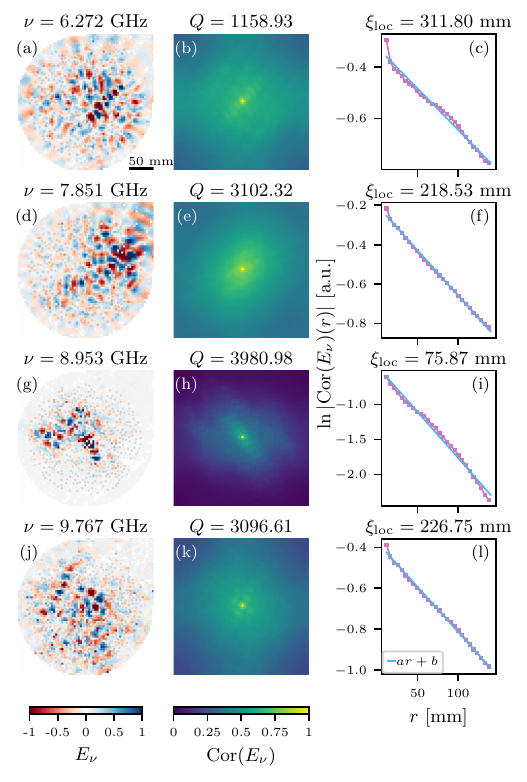}}
    \caption{
    Spatial modal structure, field amplitude spatial autocorrelation and radial profile of representative eigenmodes in a DS.
    Amplitude maps are normalized such that $\max(\abs{E_\nu})=1$, thus $\max[\mathrm{Cor}(E_{\nu})]=1$.
    Radial decays (pink dots) are obtained by performing an angular average in the autocorrelation space ($r$ is measured from the well defined autocorrelation center).
    }
    \label{fig:Dis_states}
\end{figure}

Figures~\ref{fig:Dis_states}[(a), (d), (g) and (j)] show four eigenmodes found by means of the harmonic inversion/clustering algorithms in the DS.
Since disordered systems lack of a center (in contrast to Vogel spirals), their eigenmodes are not centered in the system.
This makes the field amplitude spatial autocorrelation [see Fig.~\ref{fig:Dis_states}(b), (e), (h) and (k)], defined as
\begin{equation}
    \mathrm{Cor}(E_{\nu})=\left|\mathcal{F}^{-1}\left\{\left|\mathcal{F}\{E_{\nu}\}\right|^2\right\}\right|,
\end{equation}
the most suitable quantity to study the radial decay [see Fig.~\ref{fig:Dis_states}(c), (f), (i) and (l)]~\cite{Laurent_2007}. 
As it can be observed, all eigenmodes present large quality factors $Q$ and a clear exponential radial decay of their autocorrelation functions no matter their spatial extension, frequency, the position of their center, or the amplitude distribution with respect to the center of the mode, as expected.
The localization lenght $\xi_{\mathrm{loc}}$ is extracted by assuming $\mathrm{Cor}(E_{\nu})\propto \exp\left(-r/\xi_{\mathrm{loc}}\right)$.

\begin{figure}
    \centering
    \includegraphics{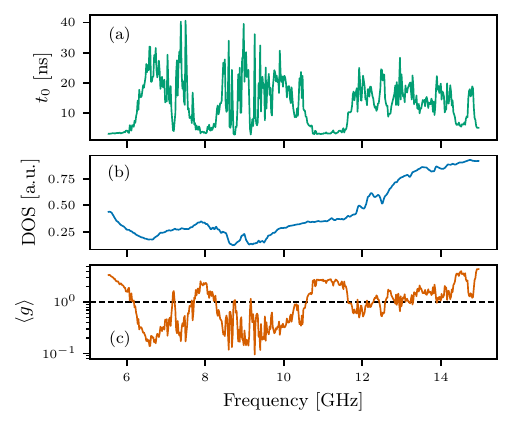}
    \caption{
    (a) Characteristic decay time $t_0$ as a function of the frequency. The total frequency range have been mapped by 472 frequency filters spaced by $\Delta f=0.02$~GHz with $\Delta\omega=0.01$~GHz.
    (b) Experimental density of states (DOS).
    (c) Experimental Thouless conductance $\expval{g}$ as a function of the frequency. The dotted line indicates $\expval{g}=1$.
    }
    \label{fig:Dis_transport}
\end{figure}

The characteristic time $t_0$, the DOS and the average Thouless conductance $\expval{g}$ are computed and plotted in Fig.~\ref{fig:Dis_transport}[(a), (b) and (c), respectively].
Similarly to the GA spiral case, $t_0$ varies from flat valleys to high peaks. Nevertheless, in contrast to the GA spiral case, no band gap can be clearly observed in the DOS.
The maxima $t_0$-values found in the GA spiral case are $t_{0,M}=40.84$~ns, $t_{0,M}=33.02$~ns and $t_{0,M}=30.11$~ns for the first and second confined regions, respectively, while in the disordered case $t_{0,M}=40.58$~ns.
Hence we conclude that the characteristic energy decay time in disordered and aperiodic Vogel spiral structures is of the same order.
Finally, the average Thouless conductance shows a fast decay of around one order of magnitude in the frequency windows where exponential eigenmodes are found by the harmonic inversion/clustering method.

\bibliography{references}

\end{document}